\documentclass[12pt]{article}
\usepackage{amssymb, amsfonts, graphicx, hyperref, psfrag, epsfig, mathrsfs}
\newcommand{\be}{\begin{equation}}
\newcommand{\bea}{\begin{eqnarray}}
\newcommand{\eea}{\end{eqnarray}}
\newcommand{\ba}{\begin{array}}
\newcommand{\ea}{\end{array}}
\newcommand{\ee}{\end{equation}}
\newcommand{\pt}{\partial}

\csname @addtoreset\endcsname{equation}{section}

\begin{document} 
\begin{titlepage}

\title{\bf {Holographic phase transition in a \\ }{non-critical holographic model}\vspace{18pt}}
\vskip.3in
 
\author{\normalsize Sheng-liang Cui,$^1$ Yi-hong Gao,$^1$ and  Wei-shui Xu$^{2,3}$  \vspace{12pt}\\
$^1${\it\small Key Laboratory of Frontiers in Theoretical Physics,}\\
{\it\small Institute of Theoretical Physics, Chinese Academy of Science}\\ 
{\it\small P.O. Box 2735, Beijing 100190, China}\\
$^2${\it\small Department of Physics, Hanyang University, Seoul 133-791, Korea}\\
$^3${\it\small Center for Quantum Spacetime, Sogang University, Seoul 121-742, Korea}\\
{\small E-mail: { \it shlcui, gaoyh@itp.ac.cn,  wsxuitp@gmail.com}}
}

\date{}
\maketitle

\voffset -.2in \vskip 2cm \centerline{\bf Abstract} \vskip .4cm

We consider a holographic model constructed from the intersecting brane
configuration D4-$\overline{\rm{D4}}$/D4 in noncritical string theory. We study
the chiral phase diagram of this holographic QCD-like model with a finite baryon
chemical potential through the supergravity dual approximation.

\vskip 4.0cm \noindent November 2009 \thispagestyle{empty}
\end{titlepage}

\newpage

\section{ Introduction}
The AdS/CFT correspondence \cite{MaldacenaRE}-\cite{WittenZW} and
\cite{AharonyTI} is one realization of the holographic principle \cite{'t
  Hooft:1993gx}. It means that the IIB string theory on the $AdS_5\times S^5$
background is equivalent to the $\mathcal{N}=4$ supersymmetric Yang-Mills theory
on the four-dimensional boundary. By adding some flavor branes into the D3
branes background, one can introduce the fundamental flavors into the low energy
effective theorie on the intersecting region of brane configuration
\cite{KarchSH}. Therefore, some more realistic effective theories can be
constructed from string theory. The strong coupling physics in these effective
boundary theories can be investigated by their classical supergravity
duals. Recently, some holographic models and the corresponding holographic phase
transitions are studied through the gauge/gravity correspondence in
\cite{KruczenskiBE}-\cite{ParnachevBC}. And one can see the reviews
\cite{MateosAY} for more related work.

Similar to the effective theories constructed from the brane configurations in
critical string theory, one can also build some holographic models from the
intersecting brane configurations in noncritical string theory. Recently, it is
investigated in \cite{KupersteinYK}-\cite{KlebanovYA}. In \cite{MazuTP}, the
authors consider a D4-$\overline{\rm{D4}}$/D4 brane configuration. The
low-energy effective theory on the intersecting region is a QCD-like
theory. Through the dual supergravity investigation, the D4-$\overline{\rm D4}$
pairs solution connected through a wormhole is preferred in the low temperature
phase. It means the flavor branes induced chiral symmetry in guage theory will
be broken. After a confinement/deconfinement phase transition into the high
temperature phase, there exists a first order chiral phase transition at a
critical temperature in this high temperature phase. Below this temperature, the
chiral symmetry is broken. However, this symmetry will be restored above it. All
these results about the chiral symmetry breaking are similar to some holographic
models being constructed in critical string theory \cite{AharonyDA}. In this
brane configuration, the six-dimensional gravity background is the near horizon
geometry of the color D4 brane. Compared with the D-brane gravity background in
critical string theory, it has some good points since there is not compact
sphere. But the fault is that such background is not very reliable in the
gauge/gravity correspondence. The reason is now the 't Hooft coupling constant
and scalar curvature are almost of order one for these gravit backgrounds. So we
can't choose a large 't Hooft limit in this holographic model.

In this paper, we extend to study the chiral phase structure of this holographic
model with a chemical potential by using the methods in
\cite{HorigomeXU}-\cite{MatsuuraZX}. The chemical potential in the boundary
gauge theory corresponds to turn on a zero-component of gauge field on the
flavor brane in the holographic brane configurations. We find, with a chemical
potential, that a chiral symmetry breaking solution is preferred in the low
temperature phase. This result is same as the case without a chemical
potential. However, at high temperature, the chiral phase diagram is different
with the no chemical potential case. And now the phase structure depends on some
parameters, which are the chemical potential and the temperature of the black
hole background.

The paper is organized as follows. In section two, we give a short introduction
to this model. Then we will investigate the holographic phase structure with a
chemical potential in sections three and four and the Appendix. In section five,
we give our conclusion.

\section{Brane configuration}
We consider the holographic model constructed by the brane configuration
D4-$\overline{\rm D4}$/D4 in \cite{MazuTP}. The coordinates of these branes are
extended as follows
\be
\begin{array}{ccccccccccccc}
 &0 &1 &2 &3 &4 &5 \\
 N_c~~\rm{D4}: &{\rm x} &{\rm x} &{\rm x} &{\rm x} &{\rm x} &{} \\
 N_f~~\rm{D4},{\rm\overline{D4}}: &{\rm x} &{\rm x} &{\rm x} &{\rm x} &{} &{\rm
 x}
\end{array}
\label{configuration}\ee The number of the color and flavor branes satisfies 
the condition $N_c\gg N_f$. In the quenching approximation, the backreaction of
the flavor branes on the color branes can be ignored. Let the coordinate
$x_4$ be compactified on a circle $S^1$, and then the adjoint fermions on the
color D4 brane with anti-periodic boundary condition on this circle will be
decoupled from the low energy effective theory. Similar to the
Sakai-Sugimoto(SS) model \cite{SakaiCN}, the low energy effective theory is
QCD-like on the four dimensional intersecting region in this brane
configuration. And this low energy theory has a global chiral symmetry
$U(N_f)_L\times U(N_f)_R$ induced by the flavor brane pairs D4-${\rm
  \overline{D4}}$.

The near horizon geometry of the color D4 branes in noncritical string theory
is expressed as \bea
&&ds^2=\left(\frac{U}{R}\right)^2(-dt^2+dx_idx_i+f(U)dx_4^2~)+
\left(\frac{R}{U}\right)^2\frac{1}{f(U)}dU^2,\nonumber\\
&&F_6=Q_c\left(\frac{U}{R}\right)^4dt\wedge dx_1\wedge dx_2\wedge
dx_3\wedge dU\wedge dx_4,\label{backgroundzero}\\
&&e^\phi=\frac{2\sqrt{2}}{\sqrt{3}Q_c},~~R^2=15/2,~~
f(U)=1-\left(\frac{U_{KK}}{U}\right)^5,\nonumber \eea where the parameter $Q_c$
is proportional to the number of color brane $N_c$. From this background, the 't
Hooft coupling constant and curvature scalar all are of order one. Because of the high
order string corrections, this background is not very good at using the
gauge/gravity correspondece. In order to avoid a singularity in this
background, the coordinate $x_4$ will be periodic with a radius \be x_4\sim
x_4+\delta x_4=x_4+\frac{4\pi R^2}{5U_{KK}}.\label{periodica}\ee And its
corresponding Kluza-Klein (KK) mass scale of this compact dimension is \be
M_{KK}=\frac{2\pi}{\delta x_4}=\frac{5U_{KK}}{2R^2}.\label{kkscale}\ee

Passing into the finite temperature phase, there exist two supergravity
backgrounds. One is that the geometry (\ref{backgroundzero}) does a Wick
rotation $t_E=it$. So it is \bea
&&ds^2=\left(\frac{U}{R}\right)^2(~dt_E^2+dx_idx_i+f(U)dx_4^2~)+
\left(\frac{R}{U}\right)^2\frac{1}{f(U)}dU^2, \nonumber \\
&&~~f(U)=1-\left(\frac{U_{KK}}{U}\right)^5,\label{background}\eea where the
Euclidean time satisfies the period $t_E\sim t_E+\beta$, and the coordinate
$x_4$ still satisfies the periodic condition (\ref{periodica}). Since the
$\beta$ is arbitrary, the temperature $1/\beta$ of this background is
undetermined.

Another finite temperature supergravity background is the black hole
case
 \bea
&&ds^2=\left(\frac{U}{R}\right)^2(~\tilde{f}(U)dt_E^2+dx_idx_i+
dx_4^2~)+\left(\frac{R}{U}\right)^2\frac{1}{\tilde{f}(U)}dU^2,\nonumber\\
&&~~~~~~\tilde{f}(U)=1-\left(\frac{U_{T}}{U}\right)^5,
\label{blackbackground}\eea where the Euclidean time satisfies 
\be t_E\sim t_E+\delta t_E=t_E+\frac{4\pi R^2}{5U_T},\label{periodicb}\ee and
the radius of the coordinate $x_4$ is arbitrary. After comparing the free energy
between the background (\ref{background}) and (\ref{blackbackground}), one can
find there exists a first order phase transition (corresponding to the
confinement/deconfinement phase transition in the boundary theory) at a critical
temperature $\beta=\delta x_4$. Below this temperature, the background
(\ref{background}) is dominated. Otherwise, the black hole background
(\ref{blackbackground}) will be dominated \cite{MazuTP}. These results are
similar to the cases in the Sakai-Sugimoto model \cite{AharonyDA}.

In the following section, we consider a $U(1)$ baryon number symmetry. Similar
to \cite{HorigomeXU}-\cite{ParnachevBC}, in order to investigate this
holographic model with a chemical potential, we only need open the zero
component of the gauge field on the worldvolume of the flavor D4 branes. In the
following, we assume that the zero component $A_0$ and the coordinate $x_4$ only
depend on the coordinate $U$, and set $\alpha'\equiv 1$. And The Abelian
effective action of the D4 branes is \be S_{D4}=-T_4\int d^5\xi~
e^{-\phi}\sqrt{-\det(~g_{MN}+2\pi\alpha' F_{MN})}+ \mu_5\int
C_5.\label{Daction}\ee By the arguments in \cite{MazuTP}, the Chern-Simons (CS)
term $\int C_5$ need to be vanished in order to make the holographic duality to
work. So we don't consider the contributions of the CS term in the following.

\section{Low temperature}
In the low temperature phase, the induced metric on the flavor D4 branes is \be
ds^2=\left(\frac{U}{R}\right)^2(~dt_E^2+dx_idx_i~)+
\left(\frac{U}{R}\right)^2\left(f(U)\left(\frac{\partial x_4}{\partial
      U}\right)^2+
  \left(\frac{R}{U}\right)^4\frac{1}{f(U)}\right)dU^2. \label{induced} \ee
Substituting this metric into the effective action (\ref{Daction}), we get the
D4 brane action as \be S\sim \int dx_4~U^5\sqrt{f(U)
  +\left(\frac{R}{U}\right)^4(f(U)^{-1}U'^2-(2\pi
  A_0')^2)},\label{inducedactiona}\ee where $U'=dU/dx_4$ and
$A'_0(U)=dA_0/dx_4$. Then the equation of motion of the coordinate $x_4(U)$ and
the zero component $A_0(U)$ are derived as \bea \frac{d}{dx_4} \left[\frac{U^5
    f(U)}{\sqrt{f(U)+\left(\frac{R}{U}\right)^4(f(U)^{-1}U'^2-(2\pi
      A_0')^2)}}\right]=0,\cr \frac{d}{dx_4}
\left[\frac{U^5\left(\frac{R}{U}\right)^42\pi
    A_0'}{\sqrt{f(U)+\left(\frac{R}{U}\right)^4(f(U)^{-1}U'^2-(2\pi
      A_0')^2)}}\right]=0. \label{eoma}\eea After doing one integration, we get
two equations \bea &&U'^2=\frac{f(U)^2\left[f(U)(U^{10}
    +C_1^2\left(\frac{U}{R}\right)^4)-f(U_0)(U_0^{10}
    +C_1^2\left(\frac{U_0}{R}\right)^4)\right]}{\left(\frac{R}{U}\right)^4f(U_0)
  \left[U_0^{10} +C_1^2\left(\frac{U_0}{R}\right)^4\right]},\cr &&~~~~(2\pi
A'_0)^2 =\frac{U^8 f(U)^2}{R^8}\frac{C_1^2}{f(U_0)(U_0^{10}+C_1^2
  \left(\frac{U_0}{R}\right)^4)},\label{solutiona}\eea where the integrating
constants $U_0$ and $C_1$ satisfy $U'|_{U=U_0}=0$ and \be
C_1=\frac{U_0^5\left(\frac{R}{U_0}\right)^42\pi
  A_0'(U_0)}{\sqrt{f(U_0)-\left(\frac{R}{U_0}\right)^4 (2\pi
    A'_0(U_0))^2}}.\label{constanta}\ee Thus, we obtained a D4-$\overline{\rm
  D4}$ pairs solution connected through a wormhole $U=U_0$. And now the chiral
symmetry $U(N_f)_L\times U(N_f)_R$ is broken to $U(N_f)_{\rm{diag}}$. After
inserting these solutions (\ref{solutiona}) into the action
(\ref{inducedactiona}), we obtain the on-shell energy of this connected
configuration \be S_1\sim \int_1^\infty dy~
\frac{y^3}{\sqrt{f(y)(1+ay^{-6})-f(1)(1+a)y^{-10}}},\label{shellactiona}\ee
where $y=U/U_0$, $y_{KK}=U_{KK}/U_0$ and $a=C_1^2/R^4U_0^6$.  For the gravity
background (\ref{background}), the coordinate $x_4$ is periodic.  The flavor
branes don't have any place to end in this background. So there doesn't exist
separated flavor D4 and ${\rm \overline{D4}}$ solution. If the coordinate $x^4$
is not periodic, and there is not $f(u)$ factor in the metric
(\ref{background}), then the separated flavor solution will be existed
\cite{AntonyanVW}. Thus, the global chiral symmetry $U(N_f)_L\times U(N_f)_R$ is
always broken to its diagonal part $U(N_f)_{\rm diag}$ at low temperature.



From the equation of motion, the asymptotic distance along the direction $x^4$
between the D4 and $\overline{\rm{D4}}$ pairs can be defined as \bea
\frac{L}{2}&=&\int_0^{L/2} dx_4=\int_{U_0}^\infty dU~\frac{1}{U'}\cr
&=&\frac{R^2}{U_0}\int_1^\infty dy~
\frac{1}{y^2f(y)}\frac{1}{\sqrt{\frac{f(y)(y^{10}+ay^4)}{f(1)(1+a)}-1}}
\label{distanceaa}\\
&=&\frac{R^2}{5U_0}\int_0^1dz~\frac{z^{1/5}\sqrt{(1+a)(1-y_{KK}^5)}}{(1-
  y_{KK}^5z)\sqrt{(1-y_{KK}^5z)(1+az^{6/5})-(1-y_{KK}^5)(1+a)z^2}}. \nonumber
\eea It depends on the chemical potential and the parameter $y_{KK}$. If
choosing $a=0$, then we find it will be reduced to the no chemical potential
case in \cite{MazuTP}.

\section{High temperature}
From the high temperature background (\ref{blackbackground}), the
induced metric on the worldvolume of the D4 flavor branes can be
obtained as \be ds^2=\left(\frac{U}{R}\right)^2
(\tilde{f}(U)dt_E^2+dx_idx_i)
+\left(\frac{U}{R}\right)^2\left(\left(\frac{\partial x_4}{\partial
U}\right)^2+\left(\frac{R}{U}\right)^4\frac{1}{\tilde{f}(U)}\right)dU^2.
\label{inducedblack}\ee The same as in the low temperature phase, the
D4 action in the background (\ref{blackbackground}) is \be S\sim
\int dx_4~ U^5 \sqrt{\tilde{f}(U)
+\left(\frac{R}{U}\right)^4(U'^2-(2\pi A_0')^2)}.
\label{inducedactionb}\ee

And the equations of motion is derived as \bea  \frac{d}{dx_4}
\left[\frac{U^5
\tilde{f}(U)}{\sqrt{\tilde{f}(U)+\left(\frac{R}{U}\right)^4(U'^2-(2\pi
A_0')^2)}}\right]=0,\cr \frac{d}{dx_4}
\left[\frac{U^5\left(\frac{R}{U}\right)^42\pi
A_0'}{\sqrt{\tilde{f}(U)+\left(\frac{R}{U}\right)^4(U'^2-(2\pi
A_0')^2)}}\right]=0.\label{eomb}\eea After integrating, we get the following two
equations \bea
U'^2=\frac{\tilde{f}(U)\left[\tilde{f}(U)(U^{10}
+C_2^2\left(\frac{U}{R}\right)^4)-\tilde{f}(U_0)(U_0^{10}
+C_2^2\left(\frac{U_0}{R}\right)^4)\right]}{\left(\frac{R}{U}\right)^4
\tilde{f}(U_0)\left[U_0^{10}
+C_2^2\left(\frac{U_0}{R}\right)^4\right]},\cr (2\pi A'_0)^2
=\frac{U^8
\tilde{f}(U)^2}{R^8}\frac{C_2^2}{\tilde{f}(U_0)(U_0^{10}+
C_2^2\left(\frac{U_0}{R}\right)^4)},\qquad\qquad
\label{solutionb}\eea where the integrating constants $U_0$ and
$C_2$ satisfy $U'|_{U=U_0}=0$ and \be
C_2=\frac{U_0^5\left(\frac{R}{U_0}\right)^42\pi
  A_0'(U_0)}{\sqrt{\tilde{f}(U_0)-\left(\frac{R}{U_0}\right)^4 (2\pi
    A'_0(U_0))^2}}.\label{constantb}\ee The same as the cases in the low
temperature phase, this solution denotes the connected configuration of the
D4-$\overline{\rm D4}$ brane pairs.

Then, after substituting this solution into the action (\ref{inducedactionb}),
we obtain the on-shell energy of this configuration \be S_3\sim \int_1^\infty
dy~
\frac{y^3\sqrt{\tilde{f}(y)}}{\sqrt{\tilde{f}(y)(1+by^{-6})-\tilde{f}(1)(1+b)y^{-10}}},\label{shellactionc}\ee
where $y=U/U_0$, $y_{T}=U_T/U_0$ and $b=C_2^2/R^4U_0^6$.

And the separated solution of the D4-$\overline{\rm D4}$ brane pairs satisfies
\be\frac{dx_4}{dU}=0,~~~~~(2\pi)^2\left(\frac{dA_0}{dU}\right)^2 =
\frac{C_2^2}{C_2^2+U^{10}\left(\frac{R}{U}\right)^4}.\label{separatedb}\ee
If $U\rightarrow \infty$, then \be
2\pi\left(\frac{dA_0}{dU}\right)\sim
\frac{C_2}{R^2}U^{-3},\label{largeu}\ee which is similar to the
corresponding low temperature case. The on-shell
action of the solution (\ref{separatedb}) is \be S_4\sim
\int_{y_T}^\infty dy~\frac{y^3}{\sqrt{1+by^{-6}}}.
\label{shellactiond}\ee

In order to determine which solution is more preferred, one needs to
compare the on-energy of these two configurations. With the equations
(\ref{shellactionc}) and (\ref{shellactiond}), the energy
difference of these two configurations is \bea \Delta
S&=&S_3-S_4=\int_1^\infty dy~
\left(\frac{y^3\sqrt{\tilde{f}(y)}}{\sqrt{\tilde{f}(y)(1+by^{-6})
-\tilde{f}(1)(1+b)y^{-10}}} -\frac{y^3}{\sqrt{1+by^{-6}}}\right)\cr
&&\qquad\qquad-\int_{y_T}^1 dy~ \frac{y^3}{\sqrt{1+by^{-6}}}\cr
&=&\frac{1}{5}\int_0^1dz~\frac{1}{z^{9/5}}
\left(\sqrt{\frac{1-y_T^5z}{(1-y_T^5z)(1+bz^{6/5})-(1-y_T^5)(1+b)z^2}}
-\frac{1}{\sqrt{1+bz^{6/5}}}\right)\cr && \qquad\qquad
-\frac{1}{5}\int_1^{y_T^{-5}}dz~\frac{1}{z^{9/5}\sqrt{1+b z^{6/5}}}.
\label{energydifferenceb}\eea

After doing numerical calculations, we plotted the Fig. 1.
\begin{figure}[ht]
 \centering
 \includegraphics[width=0.8\textwidth]{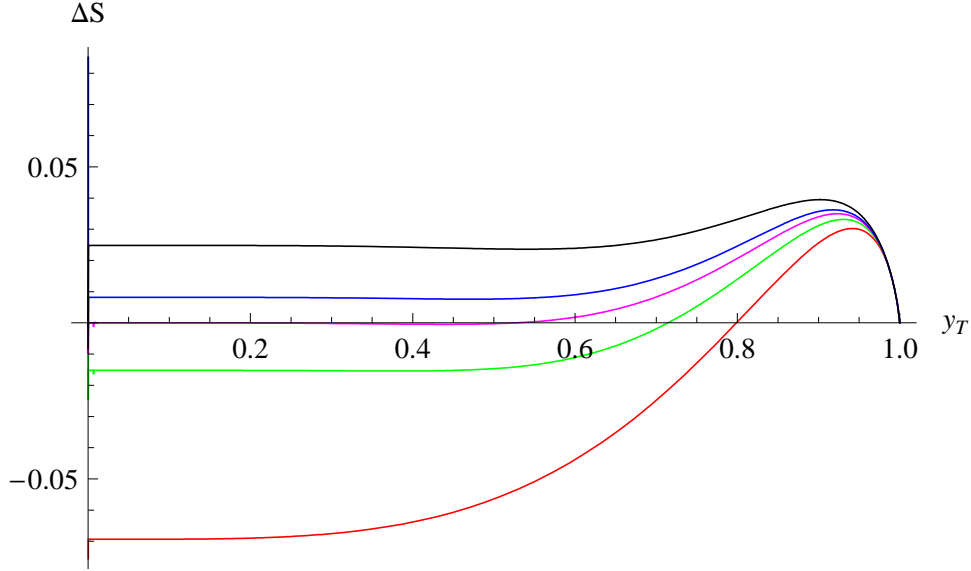}
 \caption{The energy difference $\Delta S$ varies with the
$y_T$ at various values $b=0,~ 0.10,~0.16,~0.20$ and $0.30$ (from below to above).
With increasing the parameter $b$, the energy difference $\Delta S$ is also
increased for a fixed $y_T$. }
 \label{energy}
\end{figure}
From this figure, we find there exists a turning point at $b=0.16$. If the
parameter $b$ is larger than it, then the separated configuration is dominated,
and the chiral symmetry will be always unbroken. However, if smaller than this
one, there exists two phases. One is the chiral symmetry breaking phase, the
other is the chiral symmetry restoration phase. And in the $b<0.16$ region, the
chiral phase transition point $y_T$ will become small with increasing the
parameter $b$. In all, the chiral symmetry will be broken at some smaller
parameters $b$ and $y_T$ for this holographic model. These results about the
chiral phase diagram here are similar to the cases for the Sakai-Sugimoto model
with a chemical potential \cite{HorigomeXU}.

In the background (\ref{blackbackground}), the asymptotic distance along the
direction $x^4$ between the D4 and $\overline{\rm{D4}}$ is
\bea\frac{L}{2}&=&\frac{R^2}{U_0}\int_1^\infty dy
\frac{1}{y^2}\frac{1}{\sqrt{\frac{\tilde{f}(y)(y^{10}+by^4)}{\tilde{f}(1)(1+b)}-1}}\nonumber\\
&=&\frac{R^2}{5U_0}\int_0^1dz\frac{z^{1/5}\sqrt{(1+b)(1-y_{T}^5)}}{\sqrt{(1-
    y_{T}^5z)[(1-y_{T}^5z)(1+bz^{6/5})-(1-y_{T}^5)(1+b)z^2]}}\nonumber\\
&&\qquad \equiv \frac{R^2}{5U_0} F(y_T, b). \label{distance}\eea If one chooses
$b=0$, then it will be reduced to the case without a chemical potential in
\cite{MazuTP}. And using this distance $L$, we find the temperature of the black
D4 background will be satisfied \be T=\frac{5U_T}{4\pi R^2}=\frac{y_T}{2\pi
  L}F(y_T,b).\label{temperature}\ee

Since the zero
component $A_0(y_T)=0$ at the horizon,  from the separated solution (\ref{separatedb}), 
the $A_0(U)$ satisfies \be
A_0(U)=\frac{U_0}{2\pi}\int_{y_T}^y dy~ \sqrt{\frac{b}{b+y^6}}.
\label{gauge}\ee By the gauge/gravity correspondence, the
chemical potential $\mu$ corresponds to the zero component
$A_0(U)|_{U\rightarrow \infty}$. Thus, we get \be
\mu=A_0(\infty)=\frac{R^2}{5\pi L}F(y_T, b)\int_{y_T}^\infty dy~
\sqrt{\frac{b}{b+y^6}}. \label{chemicala}\ee From this expression, we can find
the relation between the chemical potential $\mu$ and the parameters $y_T$ and
$b$. With the equation (\ref{temperature}), the chemical potential can be
expressed as \be \mu=T\frac{2R^2}{5y_T}\int_{y_T}^\infty dy~
\sqrt{\frac{b}{b+y^6}}. \ee It linearly depends on the temperature. And in
the region of the smaller values of the temperature $T$ and $b$, the chiral
symmetry is broken. However, the symmetry is restored at some larger values.

\section{Conclusions}
In this paper, we consider a holographic model constructed from the brane
configuration D4-$\overline{\rm D4}$/D4 in noncritical string theory
\cite{MazuTP}. We investigated the holographic chiral phase structure of this
noncritical string model with a finite chemical potential $\mu$. In the low
temperature phase, just as the case without a chemical potential, the chiral
symmetry breaking solution is always preferred.

In the high temperature phase, the results are different from the no chemical
potential case. Now the chiral phase structure depends on some parameters, which
are the chemical potential and temperature of the black hole background. From
the Fig. 1, one can understand very clearly on the relations of the phase
diagram with the parameters $y_T$ and $b$. Under the critical point $b=0.16$,
there exists two phases: the chiral symmetry breaking phase and the chiral
symmetry restored phase. Which system the phase lies in depends on the value of
the parameter $y_T$. And above the value $b=0.16$, the chiral symmetry will be
restored. All results of this noncritical holographic model are similar to
the corresponding cases of the SS model \cite{HorigomeXU}. These results give
some universal confirmations for some holographic models. In the appendix, we
extend to give some simple calculations about the isospin chemical
potential. One
can generalize to study the chiral phase diagram of some other holographic
models in noncritical string theory.

\subsection*{Acknowledgments}
We thank Zhu-feng Zhang for some useful discussions.

\subsection*{Appendix}
In this appendix, we generalize to consider the non-vanishing isospin chemical
potential cases by following the method in \cite{AharonyUU} and
\cite{ParnachevBC}. It corresponds to a $\rho$ meson condensation. We list the
results for the isospin chemical potential. 

For the $N_f$ coincident flavor D4 branes, the global symmetry is $U(N_f)$. Its
$U(1)$ part in the boundary corresponds to a baryon chemical potential. And the
non-Abelian part $SU(N_f)$ produces a isospin chemical potential. For
simplicity, we choose the flavor number $N_f=2$. The baryon and isospin chemical
potential are written as \be \mu_B=\frac{1}{2}(\mu_1+\mu_2),
~~~~\mu_I=\frac{1}{2}(\mu_1-\mu_2), \ee where the chemical $\mu_1$ and $\mu_2$
are the boundary value of the zero component of Abelian gauge field $A_1$ and $
A_2$ on two flavor D4-branes, respectively. If the chemical potentials satisfies
$\mu_1=\mu_2$, there only exists a baryon chemical potential, all chiral phases
is discussed above. If $\mu_1=-\mu_2$, then only the isospin chemical
potential is non-vanished. For the other case, both these chemical potentials
are existed. In the following, we mainly calculate the isospin chemical
potential in the connected D4-$\overline{\rm D4}$ brane phase with the condition
$\pt_UA_{(1)0}=-\pt_uA_{(2),0}$.

In the low temperature phase, by using the equation (\ref{solutiona}), we get
the isospin chemical potential \bea \mu_I &=&\int_{U_0}^\infty dU \pt_UA_0\cr
&=&\frac{C_1}{2\pi R^2}\int_{U_0}^\infty
\frac{U^2}{\sqrt{f(U)(U^{10}+C_1^2U^4/R^4)-f(U_0)(U_0^{10}+C_1^2U_0^4/R^4)}}.\eea
Defining $z=U_0^5/U^5$, $a=C_1^2/R^4U_0^6$ and $y_{KK}=U_{KK}/U_0$, then the
above equation becomes \be \mu_I=\frac{C_1}{10\pi R^2U_0^2}\int_0^1dz
\frac{1}{z^{3/5}\sqrt{(1-y_{KK}^5z)(1+az^{6/5})-(1-y_{KK}^5)(1+a)z^2}}.\ee

In the high temperature background, with the connected solution
(\ref{solutionb}), the isospin chemical potential is derived as \bea
\mu_I&=&\int_{U_0}^\infty dU \pt_UA_0\cr &=&\frac{C_2}{2\pi
  R^2}\int_{U_0}^\infty
\frac{U^2\tilde{f}^{1/2}}{\sqrt{\tilde{f}(U)(U^{10}+C_2^2U^4/R^4)-\tilde{f}(U_0)(U_0^{10}+C_2^2U_0^4/R^4)}}\cr
&=&\frac{C_2}{10\pi R^2U_0^2}\int_0^1dz
\frac{\sqrt{1-y_T^5z}}{z^{3/5}\sqrt{(1-y_T^5z)(1+bz^{6/5})-(1-y_T^5)(1+b)z^2}},\eea
where $z=U_0^5/U^5$, $y_T=U_T/U_0$ and $b=\frac{C_2^2}{R^4U_0^6}$.


\begin{thebibliography}{40}

\bibitem{MaldacenaRE}
  J.~M.~Maldacena,
   ``The large N limit of superconformal field theories and supergravity,''
  Adv.\ Theor.\ Math.\ Phys.\  {\bf 2}, 231 (1998)
  [Int.\ J.\ Theor.\ Phys.\  {\bf 38}, 1113 (1999)]
  [arXiv:hep-th/9711200].

\bibitem{GubserBC}
  S.~S.~Gubser, I.~R.~Klebanov and A.~M.~Polyakov,
   ``Gauge theory correlators from non-critical string theory,''
  Phys.\ Lett.\  B {\bf 428}, 105 (1998)
  [arXiv:hep-th/9802109].

\bibitem{WittenQJ}
  E.~Witten,
   ``Anti-de Sitter space and holography,''
  Adv.\ Theor.\ Math.\ Phys.\  {\bf 2}, 253 (1998)
  [arXiv:hep-th/9802150].

\bibitem{WittenZW}
  E.~Witten,
  ``Anti-de Sitter space, thermal phase transition, and confinement in  gauge
   theories,''
  Adv.\ Theor.\ Math.\ Phys.\  {\bf 2}, 505 (1998)
  [arXiv:hep-th/9803131].

\bibitem{AharonyTI}
  O.~Aharony, S.~S.~Gubser, J.~M.~Maldacena, H.~Ooguri and Y.~Oz,
   ``Large N field theories, string theory and gravity,''
  Phys.\ Rept.\  {\bf 323}, 183 (2000)
  [arXiv:hep-th/9905111].

\bibitem{'t Hooft:1993gx}
  G.~'t Hooft,
  ``Dimensional reduction in quantum gravity,''
  arXiv:gr-qc/9310026;
  L.~Susskind,
  ``The World As A Hologram,''
  J.\ Math.\ Phys.\  {\bf 36}, 6377 (1995)
  [arXiv:hep-th/9409089].

\bibitem{KarchSH}
  A.~Karch and E.~Katz,
   ``Adding flavor to AdS/CFT,''
  JHEP {\bf 0206}, 043 (2002)
  [arXiv:hep-th/0205236].

\bibitem{KruczenskiBE}
  M.~Kruczenski, D.~Mateos, R.~C.~Myers and D.~J.~Winters,
   ``Meson spectroscopy in AdS/CFT with flavour,''
  JHEP {\bf 0307}, 049 (2003)
  [arXiv:hep-th/0304032].

\bibitem{KruczenskiUQ}
  M.~Kruczenski, D.~Mateos, R.~C.~Myers and D.~J.~Winters,
  ``Towards a holographic dual of large-N(c) QCD,''
  JHEP {\bf 0405}, 041 (2004)
  [arXiv:hep-th/0311270].

\bibitem{SakaiCN}
  T.~Sakai and S.~Sugimoto,
   ``Low energy hadron physics in holographic QCD,''
  Prog.\ Theor.\ Phys.\  {\bf 113}, 843 (2005)
  [arXiv:hep-th/0412141].

\bibitem{SakaiYT}
  T.~Sakai and S.~Sugimoto,
  ``More on a holographic dual of QCD,''
  Prog.\ Theor.\ Phys.\  {\bf 114}, 1083 (2005)
  [arXiv:hep-th/0507073].

\bibitem{BabingtonVM}
  J.~Babington, J.~Erdmenger, N.~J.~Evans, Z.~Guralnik and I.~Kirsch,
   ``Chiral symmetry breaking and pions in non-supersymmetric gauge/gravity
   duals,''
  Phys.\ Rev.\  D {\bf 69}, 066007 (2004)
  [arXiv:hep-th/0306018]

\bibitem{AntonyanVW}
  E.~Antonyan, J.~A.~Harvey, S.~Jensen and D.~Kutasov,
   ``NJL and QCD from string theory,''
  arXiv:hep-th/0604017;
  E.~Antonyan, J.~A.~Harvey and D.~Kutasov,
  ``The Gross-Neveu model from string theory,''
  Nucl.\ Phys.\  B {\bf 776}, 93 (2007)
  [arXiv:hep-th/0608149];
  E.~Antonyan, J.~A.~Harvey and D.~Kutasov,
  ``Chiral symmetry breaking from intersecting D-branes,''
  Nucl.\ Phys.\  B {\bf 784}, 1 (2007)
  [arXiv:hep-th/0608177];
  Y.~h.~Gao, W.~s.~Xu and D.~f.~Zeng,
   ``NGN, QCD(2) and chiral phase transition from string theory,''
  JHEP {\bf 0608}, 018 (2006)
  [arXiv:hep-th/0605138];
Y.~h.~Gao, J.~P.~Shock, W.~s.~Xu and D.~f.~Zeng,
  ``A note on chiral symmetry breaking from intersecting branes,''
  Phys.\ Rev.\  D {\bf 76}, 046003 (2007)
  [arXiv:0704.3913 [hep-th]].

\bibitem{MateosNU}
  D.~Mateos, R.~C.~Myers and R.~M.~Thomson,
  ``Holographic phase transitions with fundamental matter,''
  Phys.\ Rev.\ Lett.\  {\bf 97}, 091601 (2006)
  [arXiv:hep-th/0605046].

\bibitem{AharonyDA}
  O.~Aharony, J.~Sonnenschein and S.~Yankielowicz,
  ``A holographic model of deconfinement and chiral symmetry restoration,''
  Annals Phys.\  {\bf 322}, 1420 (2007)
  [arXiv:hep-th/0604161];
  A.~Parnachev and D.~A.~Sahakyan,
  ``Chiral phase transition from string theory,''
  Phys.\ Rev.\ Lett.\  {\bf 97}, 111601 (2006)
  [arXiv:hep-th/0604173].

\bibitem{PeetersIU}
  K.~Peeters, J.~Sonnenschein and M.~Zamaklar,
 ``Holographic melting and related properties of mesons in a quark
  gluon
  plasma,''
  Phys.\ Rev.\  D {\bf 74}, 106008 (2006)
  [arXiv:hep-th/0606195].

\bibitem{HorigomeXU}
  N.~Horigome and Y.~Tanii,
 ``Holographic chiral phase transition with chemical potential,''
  JHEP {\bf 0701}, 072 (2007)
  [arXiv:hep-th/0608198].

\bibitem{KobayashiSB}
  S.~Kobayashi, D.~Mateos, S.~Matsuura, R.~C.~Myers and R.~M.~Thomson,
   ``Holographic phase transitions at finite baryon density,''
  JHEP {\bf 0702}, 016 (2007)
  [arXiv:hep-th/0611099].

\bibitem{MateosVC}
  D.~Mateos, S.~Matsuura, R.~C.~Myers and R.~M.~Thomson,
   ``Holographic phase transitions at finite chemical potential,''
  JHEP {\bf 0711},  085 (2007) 
  arXiv:0709.1225 [hep-th].

\bibitem{MatsuuraZX}
  S.~Matsuura,
   ``On holographic phase transitions at finite chemical potential,''
  JHEP {\bf 0711},  098 (2007)
  [arXiv: 0711.0407[hep-th]].

\bibitem{AharonyUU}
  O.~Aharony, K.~Peeters, J.~Sonnenschein and M.~Zamaklar, 
  ``Rho meson condensation at finite isospin chemical potential in a holographic model for QCD,''
   JHEP {\bf 0802}, 071 (2008) 
   [arXiv:0709.3948 [hep-th]].

\bibitem{ParnachevBC}
  A.~Parnachev,
  ``Holographic QCD with Isospin Chemical Potential,''
  JHEP {\bf 0802},  062 (2008) 
  arXiv:0708.3170 [hep-th].

\bibitem{MateosAY}
  D.~Mateos,
  ``String Theory and Quantum Chromodynamics,''
  Class.\ Quant.\ Grav.\  {\bf 24}, S713 (2007)   
  [arXiv:0709.1523 [hep-th]];
  J.~Erdmenger, N.~Evans, I.~Kirsch and E.~Threlfall,
   ``Mesons in Gauge/Gravity Duals - A Review,''
   Eur.\ Phys.\ J.\ A {\bf 35}, 81 (2008)    
  [arXiv:0711.4467 [hep-th]].

\bibitem{KupersteinYK}
  S.~Kuperstein and J.~Sonnenschein,
  ``Non-critical supergravity (d $>$ 1) and holography,''
  JHEP {\bf 0407}, 049 (2004)
  [arXiv:hep-th/0403254].

\bibitem{KupersteinYF} 
  S.~Kuperstein and J.~Sonnenschein,
  ``Non-critical, near extremal AdS(6) background as a holographic
  laboratory of four dimensional YM theory,'' JHEP {\bf 0411}, 026
  (2004) [arXiv:hep-th/0411009].

\bibitem{Bigazzi:2005md}
  F.~Bigazzi, R.~Casero, A.~L.~Cotrone, E.~Kiritsis and A.~Paredes,
  ``Non-critical holography and four-dimensional CFT's with fundamentals,''
  JHEP {\bf 0510}, 012 (2005)
  [arXiv:hep-th/0505140].

\bibitem{CaseroSE}
  R.~Casero, A.~Paredes and J.~Sonnenschein,
   ``Fundamental matter, meson spectroscopy and non-critical string / gauge
  duality,''
  JHEP {\bf 0601}, 127 (2006) 
  [arXiv:hep-th/0510110].

\bibitem{MazuTP}
  V.~Mazu and J.~Sonnenschein,
   ``Non critical holographic models of the thermal phases of QCD,''
  JHEP {\bf 0806}, 091 (2008) 
  [arXiv:0711.4273 [hep-th]].


\bibitem{CuiXU}
  S.~l.~Cui, Y.~h.~Gao, Y.~Seo, S.~j.~Sin and W.~s.~Xu,
   ``Note on a non-critical holographic model with a magnetic field,''
  arXiv:0910.2661 [hep-th]


\bibitem{KlebanovYA}
  I.~R.~Klebanov and J.~M.~Maldacena,
  ``Superconformal gauge theories and non-critical superstrings,''
  Int.\ J.\ Mod.\ Phys.\  A {\bf 19}, 5003 (2004)
  [arXiv:hep-th/0409133].

\end{thebibliography}
\end{document}